\documentstyle[12pt]{article}
\newcommand{\GeV}{~\mbox{GeV}}
\newcommand{\TeV}{~\mbox{TeV}}

\newcommand{\mod}{{\rm mod} \hspace{2mm}}

\newcommand{\vev}[1]{ \left\langle {#1} \right\rangle }
\begin{document}
\baselineskip 0.7cm

\begin{titlepage}

\begin{flushright}
UT-938\\
\end{flushright}
 
\vskip 2cm
\begin{center}
 {\large\bf Nonanomalous R-symmetry in supersymmetric unified theories 
of quarks and leptons}
 \vskip 1.2cm
 Kiichi Kurosawa,$^1$ Nobuhito Maru$^1$ and T. Yanagida$^{1,2}$

 \vskip 0.4cm

 {\it $^1$ Department of Physics, University of Tokyo,
 Tokyo 113-0033, Japan}\\
 {\it $^2$ Research Center for the Early Universe, University of Tokyo,
 Tokyo 113-0033, Japan}
\vskip 0.2cm
\vskip 2cm
\abstract{
 A discrete R-symmetry often appears as an exact gauge symmetry in the
 low energy effective theory of superstring theories. We search for such
 discrete R-symmetries from a phenomenological point of view and find
 that $Z_{9R}$ and $Z_{18R}$ are candidates of the nonanomalous
 R-symmetry in the case of the minimal supersymmetric standard model. 
 We also find $Z_{4R}$ and $Z_{20R}$ in the case that quarks and leptons
 are embedded in the SU(5) GUT multiplets. Interesting is that in the
 latter case all the solutions predict some extra matter multiplets 
 and we find that the simplest choice of the extra matters is to take a
 pair of  ${\bf 5}$ and ${\bf 5^*}$ of SU(5)$_{\rm GUT}$ whose mass is
 of order the SUSY breaking scale $\sim 1\TeV$. We  emphasize that the
 presence of such  extra matters is testable in future hadron collider
 experiments.
}
\end{center}
\end{titlepage}

%
\section{Introduction}
%
Higher dimensional supergravity theories such as superstring theories
always contain R-symmetry, which is naturally broken down to its
discrete subgroup $Z_{NR}$ by the compactification of the extra space
\cite{string,IWY}. This discrete R-symmetry should be nonanomalous
since it is a gauge symmetry. Thus, the presence of a nonanomalous
discrete $Z_{NR}$ is one of important low-energy ingredients of higher
dimensional supergravity. On the other hand, the discrete R-symmetry is
a crucial symmetry that suppresses the dangerous dimension-four
operators \cite{P_decay, dim_5} causing too rapid proton decays 
and hence it guarantees the matter stability in the supersymmetric 
standard model (SUSY SM). 
This discrete R-symmetry is also very important to understand the light
Higgs doublet in the standard model. 
That is, the discrete R-symmetry may forbid the 
SUSY-invariant mass term ($\mu$-term) of the Higgs chiral multiplets. 
Furthermore, when the R-symmetry breaking is linked to
the SUSY breaking the Higgs chiral multiplets acquire a
SUSY-invariant mass of order the gravitino mass 
$m_{3/2}$ $\sim 1$ TeV \cite{masiero}.\footnote{
Throughout this paper we assume that the SUSY breaking is mediated to 
the MSSM sector by nonrenormalizable interactions suppressed by
the Planck scale.}

Our question is whether such a nonanomalous discrete R-symmetry exists or
not in a phenomenological point of view.\footnote{
We mean by ``nonanomalous'' that anomalies are cancelled out among only
fermions. In this paper we do not pursue a possibility of anomaly
cancellations by the Green-Schwartz mechanism \cite{GS}.
} 
We show in this paper that it indeed exists in the minimal SUSY
standard model (MSSM) and that the nonanomalous discrete
R-symmetry is restricted to either $Z_{9R}$ or $Z_{18R}$.\footnote{
A related topic is discussed by Ibanez and Ross in Refs.\cite{IR}.
Our point is that we can say more about the discrete R-symmetry
by focusing on the above relation between the SUSY breaking
and the $\mu$-term of the Higgs multiplets. 
} 
However, we point out that the discrete R-symmetry is always anomalous 
if we take the R-charge assignment consistent with 
the grand unification model (GUT) of quarks and leptons. 
We also find that the anomalies are cancelled out 
by a pair of extra matter multiplets ${\bf 5}$ and ${\bf 5^*}$ 
of SU(5)$_{\rm GUT}$ at the SUSY-breaking scale $\sim 1$ TeV,
and that the allowed R-symmetries are $Z_{4R}$ or $Z_{20R}$.
We explicitly construct a unification model based on a semi-simple gauge
group SU(5)$_{\rm GUT}\times$U(3)$_{\rm H}$ \cite{yanagida,hy,iy}. 

\section{Discrete R-symmetry in MSSM}
We first discuss conditions that an anomaly-free discrete R-symmetry
$Z_{NR}$ should satisfy in the MSSM (including right-handed neutrinos).
Then we show that such an R-symmetry does exist and that it is
restricted to either $Z_{9R}$ or $Z_{18R}$. We also discuss their
phenomenological aspects.

\begin{table}
 \begin{center}
  {\renewcommand{\arraystretch}{1.1}
  \begin{tabular}{|c||c|c|c|c|c|c|c|c||c|}
   \hline 
   & \makebox[8mm]{$Q$} & \makebox[8mm]{$\bar{U}$} & 
   \makebox[8mm]{$\bar{D}$} & \makebox[8mm]{$L$} &
   \makebox[8mm]{$\bar{E}$} & \makebox[8mm]{$\bar{N}$} &
   \makebox[8mm]{$H$} & \makebox[8mm]{$\bar{H}$} & 
   \makebox[8mm]{$\theta$} \\
   \hline
   SU(3)$_C$ & ${\bf 3}$ & ${\bf 3^*}$ & ${\bf 3^*}$ & 
   ${\bf 1}$ & ${\bf 1}$ & ${\bf 1}$ & ${\bf 1}$ & ${\bf 1}$ & \\
   \hline
   SU(2)$_L$ & ${\bf 2}$ & ${\bf 1}$ & ${\bf 1}$ & 
   ${\bf 2}$ & ${\bf 1}$ & ${\bf 1}$ & ${\bf 2}$ & ${\bf 2}$ & \\
   \hline
   U(1)$_Y$ & $1/6$ & $-2/3$ & $1/3$ & 
   $-1/2$ & $1$ & $0$ & $1/2$ & $-1/2$ & \\
   \hline
   $Z_{NR}$ & $q$ & $\bar{u}$ & $\bar{d}$ & 
   $l$ & $\bar{e}$ & $\bar{n}$ & $h$ & $\bar{h}$ & $\alpha$ \\ 
   \hline
  \end{tabular}
  }
  \caption{The matter content of the MSSM. $\theta$ is the Grassmann
  coordinate. All $Z_{NR}$ charges are taken to be integer. If one
  chooses the $Z_{NR}$ charge of the Grassmann coordinate $\theta$ to be
  unity as done in most of literatures, the $Z_{NR}$ charges of the
  chiral multiplets should be taken fractional. We consider only a
  generation independent $Z_{NR}$ symmetry in this paper, so generation
  indices are omitted.}
  \label{MSSM}
 \end{center}
\end{table}
The MSSM matter content and the $Z_{NR}$ charges are listed in the
Table \ref{MSSM}.
It is assumed through this paper that tiny neutrino masses are
generated by the see-saw mechanism \cite{see-saw}.
Then the superpotential in the MSSM is 
\begin{equation}
\label{yukawa}
W = Q\bar{U}H + Q\bar{D}H + L\bar{E}\bar{H} + L\bar{N}\bar{H} 
+ \frac{1}{2}M_m \bar{N}\bar{N}+\mu H\bar{H}, 
\end{equation}
where $M_m$ is a Majorana mass for the right-handed neutrinos and
$\mu$ is a SUSY-invariant mass for the Higgs doublets.
Since the superpotential has the $Z_{NR}$ charge $2\alpha$
it imposes the following conditions for 
the $Z_{NR}$ charges,\footnote{
Here we have assumed that the $Z_{NR}$ charge of the Majorana mass 
$M_m$ is zero, that is, the generation of the Majorana mass scale is
independent of the SUSY breaking.
} 
\begin{eqnarray}
\label{up}
q + \bar{u} + h &=& 2\alpha \quad \mod N, \\
\label{down}
q + \bar{d} + \bar{h} &=& 2\alpha \quad \mod N, \\
\label{lepton}
l + \bar{e} + \bar{h} &=& 2\alpha \quad \mod N, \\
\label{neutrino}
l + \bar{n} + h &=& 2\alpha \quad \mod N, \\
\label{majorana}
2 \bar{n} &=& 2\alpha \quad \mod N, 
\end{eqnarray}
As for the last term in the superpotential, $\mu$
must be of the order of the weak scale. One of the most 
attractive mechanisms to generate the $\mu$-term is the Giudice-Masiero
mechanism \cite{masiero}. That is, the K\"ahler potential,
\begin{eqnarray}
 K&=&\kappa H \bar{H},
\end{eqnarray}
induces the $\mu$-term of the weak scale,
\begin{eqnarray}
 W&=& \kappa\vev{W} H \bar{H},
\end{eqnarray} 
where $\kappa$ is ${\cal O}(1)$ constant and $\vev{W} = m_{3/2}
\sim 1\TeV$.
Hereafter, we take the Planck scale $M_P = 2.4\times 10^{18}\GeV$
to be unity. 
In order to obtain the $\mu$-term through the Giudice-Masiero
mechanism the following conditions have to be imposed: 
\begin{eqnarray}
\label{Masiero}
h + \bar{h} &=& 0 \quad \mod N, \\
\label{Masiero2}
h + \bar{h} &\ne& 2\alpha \quad \mod N.
\end{eqnarray}
Eq.(\ref{Masiero2}) is necessary to prevent the Higgs doublets 
from having a Planck scale mass. It can be rewritten as
\begin{eqnarray}
 2\alpha &\ne& 0  \quad \mod N,
  \label{Masiero2'}
\end{eqnarray}
from Eq. (\ref{Masiero}). This condition is also desirable in viewpoint
of the cosmological constant problem since $\vev{W}=0$ as long as the
$Z_{NR}$ symmetry is unbroken. Otherwise, we have $\vev{W}={\cal O}(1)$ 
leading to a negative cosmological constant of the order of the
Planck scale \cite{iy}. Furthermore, this condition excludes
the $N=2$ case including the R-parity, so that $N \ge 3$.

The anomaly cancellation conditions 
we should consider \cite{IR} are\footnote{
The mixed anomaly $Z_{NR}[$U(1)$_Y]^2$ does not
give any useful condition because the U(1)$_Y$ charge is not
quantized \cite{IR,BD}.  
} 
\begin{eqnarray}
\label{su3}
{\cal A}_3 &=& \frac{3}{2} 
\left\{ 2(q - \alpha) + (\bar{u} - \alpha) 
+ (\bar{d} - \alpha) \right\} + 3\alpha = 
\frac{N}{2} k, \\
\label{su2}
{\cal A}_2 &=& \frac{3}{2} 
\left\{ 3(q - \alpha) + (l - \alpha) \right\} + 
\frac{1}{2} \left\{(h -\alpha) + (\bar{h} - \alpha) \right\} 
+ 2\alpha = \frac{N}{2} k',
\end{eqnarray}
where ${\cal A}_3$ and ${\cal A}_2$ are anomaly coefficients for 
$Z_{NR}[$SU(3)$_C]^2$ and $Z_{NR}[$SU(2)$_L]^2$, respectively. 
$k$ and $k'$ are integers.
These conditions are simplified to 
\begin{eqnarray}
\label{su3'}
2q + \bar{u} + \bar{d} &=& 2\alpha + \frac{N}{3}k, \\
\label{su2'}
3(3q + l) + h + \bar{h} &=& 10\alpha + Nk'.
\end{eqnarray}
Using Eqs.(\ref{up}), (\ref{down}) and (\ref{Masiero}) we obtain
\begin{equation}
\label{ud}
2q + \bar{u} + \bar{d} = 4\alpha \quad \mod N.
\end{equation}
Then the anomaly cancellation condition for $Z_{NR}$ [SU(3)$_C$]$^2$
in Eq.(\ref{su3}) becomes
\begin{equation}
\label{selection}
N = \frac{6\alpha}{k}.
\end{equation}
Together with Eqs.(\ref{neutrino})-(\ref{Masiero}), on the other hand,
the anomaly cancellation condition for
$Z_{NR}$ [SU(2)$_L$]$^2$ becomes 
\begin{equation}
\label{su2''}
9q - 3h - 7\alpha = Nk' + \eta \frac{N}{2}k'',
\end{equation}
where  $k''$ is an integer and $\eta=0$ or $1$ for $N$ being odd or even. 
Note that the integer $k$ should not be a multiple of three from
Eqs.(\ref{Masiero2'}) and (\ref{selection}). 
Then $N$ must be a multiple of three. 
It, however, imposes that the integer $\alpha$ has to be a multiple of
three from Eq.(\ref{su2''}). 
At last it is found that $N=9M$, where $M$ is an integer.

Finally, we discuss the other anomaly cancellation conditions.
The cancellation condition of the mixed gravitational anomaly
\cite{IR} is given as follows:
\begin{eqnarray}
\label{zn}
{\cal A}_{\rm gravity} 
&=& 18(q-\alpha) + 9(\bar{u}-\alpha) + 9(\bar{d}-\alpha) 
+ 6(l-\alpha) + 3(\bar{e}-\alpha) +3(\bar{n}-\alpha) \nonumber \\
&&+ 2(h-\alpha) + 2(\bar{h}-\alpha) 
+ (8 + 3 + 1 - 21) \alpha \nonumber \\
&=& -13 \alpha = mN + \eta n\frac{N}{2}, 
\end{eqnarray}
where $m, n$ are integers.
For $N=$ odd case, 
\begin{equation}
\label{odd}
{\cal A}_{\rm gravity}
= 5 \alpha \quad {\rm mod}\hspace{1mm}\frac{6\alpha}{k}. 
\end{equation}
For $N=$ even case, 
\begin{equation}
\label{even}
{\cal A}_{\rm gravity}
= 2 \alpha \quad {\rm mod}\hspace{1mm}\frac{3\alpha}{k}. 
\end{equation}
One can see that the MSSM matter content does not cancel 
the mixed gravitational anomaly. 
However, it is easy to cancel this anomaly 
by introducing five (two) singlets with vanishing R-charges 
for $N=$ odd (even), respectively.\footnote{
There is another possibility to cancel 
the mixed gravitational anomaly. 
This is done by introducing only one singlet with R-charge $2\alpha$ 
for $N=$ odd and even cases. 
In this case, this singlet remains massless.} 
These singlets have masses of the order of the gravitino mass through
the Giudice-Masiero mechanism like the Higgs doublets. 
The cubic anomaly $Z_{N}^3$ quite depends on 
the charges of heavy fields. 
Thus, it is not so useful to 
constrain the low energy physics 
as long as we have no information about 
the massive sector \cite{IR,BD}. 
Indeed, we have checked that 
the cubic anomalies are not cancelled 
by only the MSSM matter content (and the above mentioned singlets) 
but cancelled 
by adding some extra heavy fields with fractional charges. 

To summarize, we have above shown that the discrete $Z_{NR}$ symmetry
should satisfy the following eight conditions:
four  from the Yukawa couplngs (Eqs.(\ref{up})-(\ref{neutrino})),
one from the Majorana mass for the right-handed neutrinos 
(Eq.(\ref{majorana})), 
one from the Giudice-Masiero mechanism (Eq.(\ref{Masiero})),
and two from the anomaly cancellation conditions
(Eqs.(\ref{su3}) and (\ref{su2})).
Moreover, note that there is one constraint (Eq.(\ref{Masiero2'})) 
from the Giudice-Masiero mechanism. 

Before solving the above eight conditions,
we show that the number of independent $Z_{NR}$ charge assignments
is finite. Since there are ten free parameters (nine charges given in
Table \ref{MSSM} and one parameter $N$),
it seems that there are two free parameter and that
the number of allowed charge assignments is infinite.
However, these assignments are not independent of each other. 
First, the following shift of a $Z_{NR}$ charge assignment by the
integer hypercharge (the hypercharge multiplied by six)
gives a physically equivalent charge assignment 
because of the presence of the U(1)$_Y$ gauge symmetry \cite{IR}:
\begin{eqnarray}
{\cal Q} & \to & {\cal Q}+(1,-4,2,-3,6,0,3,-3,0).
\label{shift}\\
&& ({\cal Q} \equiv 
(q,\bar{u},\bar{d},l,\bar{e},\bar{n},h,\bar{h},\alpha))
\end{eqnarray}
Second, there is a freedom of total normalization of the $Z_{NR}$
charges. Consequently,  
the number of independent charge assignments is finite and
indeed only twelve as will be shown below.

Now we present the explicit $Z_{NR}$ charge assignments.
As in Ref.\cite{IR},
it is convenient to represent the $Z_{NR}$ charge assignments 
using a linear combination of some independent generators such as
the lepton number. 
First we use five of eight conditions, that is, four from the
Yukawa couplings and one from the Giudice-Masiero mechanism.
In addition, we set the charge $q$ equal to $\alpha$ using the  
shift given in Eq.(\ref{shift}).
\begin{table}
 \begin{center}
  \begin{tabular}{|c|rrrrrrrrr|}
   \hline \raisebox{0ex}[12pt]{}
   \makebox[8mm]{} & \makebox[5mm]{\hfill $q$} &
   \makebox[5mm]{\hfill $\bar{u}$} & \makebox[5mm]{\hfill $\bar{d}$} &
   \makebox[5mm]{\hfill $l$} & \makebox[5mm]{\hfill $\bar{e}$} &
   \makebox[5mm]{\hfill $\bar{n}$} & \makebox[5mm]{\hfill $h$} &
   \makebox[5mm]{\hfill $\bar{h}$} & \makebox[5mm]{\hfill $\alpha$} \\ 
   \hline
   $P$ & $1$ &  $1$ & $1$ &  $1$ & $1$ &  $1$ & $0$ &  $0$ & $1$ \\
   $R$ & $0$ & $-1$ & $1$ &  $0$ & $1$ & $-1$ & $1$ & $-1$ & $0$ \\
   $L$ & $0$ &  $0$ & $0$ & $-1$ & $1$ &  $1$ & $0$ &  $0$ & $0$ \\
   \hline
  \end{tabular}
  \caption{The charges of the three generators.
  $P$, $R$ and $L$ correspond to the ${\cal R}$-symmetry 
  (the generalized R-parity),
  the right-handed isospin and the lepton number, respectively.}
  \label{generator}
 \end{center}
\end{table}
Since three charges remain undetermined,
we choose the ${\cal R}$-symmetry, the right-handed isospin and the 
lepton number given in Table \ref{generator} as three independent
generators. 
Then the $Z_{NR}$ charge assignments are represented by a linear
combination of these three generators.
\begin{eqnarray}
 Z_{NR} : P^\alpha R^\beta L^\gamma \quad
(\alpha,\beta,\gamma = 0,1,\cdots, N-1).
\end{eqnarray}
This expression means that the $Z_{NR}$ charges are 
assigned as ${\cal Q}=\alpha P+\beta R+\gamma L $.
The three integers $\alpha$, $\beta$ and $\gamma$ are determined by the
remaining three conditions and one constraint as follows:
\begin{eqnarray}
6\alpha&=&0 \quad \mod N, \\
 -2\beta+2\gamma &=&0 \quad \mod N, \\ 
  2\alpha-3\gamma&=&0 \quad \mod N, \\
2\alpha&\neq& 0 \quad \mod N. 
\end{eqnarray}
Solving these four equations and removing ambiguity of total
normalization of the charges, we find that $N=9$ or $18$ and obtain
the following twelve charge assignments:
\begin{eqnarray}
 Z_{9R} &:& P^3 (RL)^{3m+2},
\label{Z9R}\\
 Z_{18R} &:& P^3(RL)^{6m+2},\ \ 
             P^3(RL)^{6m+2} R^9,\ \ 
             P^6(RL)^{6m+4} R^9,
\label{Z18R}
\end{eqnarray}
where $m=0,1,2$. Implications of these charge assignments are understood
as follows. Recall the equivalence of the shift
by the hypercharge $Y$. That is, the generator $RL$ is equivalent to
the baryon number $B = RLY^{-2}$, and the generator $R^9$ is 
the matter parity $(RY^{-6})^9$. 
Therefore, the above $Z_{NR}$ charge assignments are understood as the
combination of the ${\cal R}$-symmetry and the baryon number (plus the
matter parity for $Z_{18R}$).\footnote{
The lepton number and the Peccei-Quinn symmetry are violated
by the Majorana mass term and the Giudice-Masiero mechanism,
respectively. This is why the above charge assignments are given by 
combinations of the $R$-symmetry and the baryon number.
}
And the patterns of these combinations are determined by the anomaly
cancellation conditions.

One can easily check that these discrete R-symmetries forbid 
all of the dimension four and five operators 
which violate the baryon number or the lepton number 
except for the operator $LLHH$.
In the case of $Z_{9R}$, for example, an allowed operator has to
satisfy the following condition:
\begin{eqnarray}
 3 Q_P +(3m+2) Q_B &=& 6 \quad \mod N(=9),
  \label{violation}
\end{eqnarray}
where $Q_P$ is the ${\cal R}$-charge of the operator
($Q_P=1$ for the Grassmann coordinate $\theta$) and $Q_B$
the baryon number. 
Obviously, the unwanted dimension four and five operators 
cannot satisfy the above condition.  

When SUSY is broken, the $Z_{9R}$ and $Z_{18R}$ symmetries are
spontaneously broken by the condensation of the superpotential
$\vev{W}\neq 0$.
The superpotential has a $Z_{NR}$ charge $2\alpha = 6$ ($Z_{9R}$ and
the first two of $Z_{18R}$'s) or $2\alpha=12$ (the last one of
$Z_{18R}$'s).
Therefore, the $Z_{9R}$ and $Z_{18R}$ symmetries are broken down as
follows: 
\begin{eqnarray}
  Z_{9R} \to Z_{3} &:&  (RL)^2 \\
  Z_{18R} \to Z_{6R}  &:&  P^3(RL)^2, P^3 (RL)^2 R^3 \\
           Z_{6}  &:&  (RL)^4 R^3
\end{eqnarray}
Although the order parameter of the SUSY breaking is 
$\vev{W} \sim 10^{-15}$, 
it might bring about significant cosmological signatures
through the effective dimension four operators.
In the cases of Z$_{9R}$ and the last of Z$_{18R}$'s there
remains no discrete R-symmetry. Then the lepton number violating
dimension four operators
($Q \bar{D} L$, $LL\bar{E}$) appear with a coupling 
$\lambda \sim \vev{W} \sim 10^{-15}$, 
so that the lightest SUSY particle (LSP) is unstable.\footnote{
The baryon number violating one ($\bar{U} \bar{D} \bar{D}$)
is still forbidden, so that there are no proton decays through
the dimension four operators.
} 
If the LSP is the photino as in most gravity-mediated SUSY breaking
models, its lifetime \cite{Dawson} is 
\begin{eqnarray}
  \tau_{LSP} \sim \left(\frac{\alpha\lambda^2}{128\pi^2}
             \frac{m_{LSP}^5}{m_{\tilde f}^4} \right)^{-1} 
            \sim \left(\frac{\lambda}{10^{-15}}\right)^{-2}
                  \times 10^9\ s,
\end{eqnarray}
where $m_{\tilde f}$ is the related squark or slepton mass  and
in the last equality we have assumed that 
$m_{LSP} \sim m_{\tilde f} \sim 100\GeV$.
The LSP with a life time of order $10^9 s$ is excluded because
it causes a significant distortion of cosmic microwave background or 
spoils the success of the big bang nucleosynthesis \cite{decay}.
Even if the coupling $\lambda$ are further suppressed 
and the lifetime of the LSP
is longer than the age of the universe, the absence of a high
energy neutrino background requires that $\tau_{LSP} > 10^{17}yr$
\cite{HEN}, which imposes 
an unnatural tuning on the coupling $\lambda < 10^{-23}$.

Consequently, the first two charge assignments of $Z_{18R}$ in
Eq.(\ref{Z18R}) are phenomenologically viable. 
In these cases a discrete $R$ symmetry $Z_{6R}$ remains unbroken 
and then the unwanted dimension four operators 
are absent even after the SUSY breaking. 

\section{GUT and discrete R-symmetry}
In this section, we extend the previous analysis to the case 
that the $Z_{NR}$ charge assignment is consistent with 
the grand unification model (GUT) of quarks and leptons. 
That is, we assign the same $Z_{NR}$ charges to the quarks and leptons
in the same multiplet of SU(5)$_{\rm GUT}$ as follows:  
\begin{eqnarray}
 t &\equiv& q=\bar{u}=\bar{e},\\
 \bar{f} &\equiv& \bar{d}=l.
\end{eqnarray}
Then the conditions Eqs.(\ref{up})-(\ref{majorana}) are rewritten as 
\begin{eqnarray}
 2t + h &=& 2\alpha \quad \mod N, 
  \label{yukawa-tth}\\
 t+ \bar{f} + \bar{h} &=& 2\alpha \quad \mod N, 
  \label{yukawa-tfh}\\
 \bar{f} + \bar{n} + h &=& 2\alpha \quad \mod N, 
  \label{yukawa-fnh}\\
 2 \bar{n} &=& 2\alpha \quad \mod N. 
  \label{majorana2}
\end{eqnarray} 
In addition, Eqs.(\ref{Masiero}) and (\ref{Masiero2})
have to be imposed for the Giudice-Masiero mechanism to work.
In the SU(5)$_{\rm GUT}$ respected case
the anomaly cancellation conditions Eqs. (\ref{su3}) and (\ref{su2}) 
are obtained as
\begin{eqnarray}
 {\cal A}_3 &=& 3\alpha\ =\ \frac{N}{2}k,
  \label{A3}\\
 {\cal A}_2 &=& \alpha\ =\ \frac{N}{2}k',
  \label{A2}
\end{eqnarray}
from Eqs.(\ref{Masiero}), (\ref{yukawa-tth}) and (\ref{yukawa-tfh}).

Notice that Eq.(\ref{A2}) contradicts Eq.(\ref{Masiero2'}).
This contradiction implies that there are some extra matter multiplets
contributing to the anomalies than the MSSM particles. 
Since these extra matter multiplets must form complete multiplets of
the SU(5)$_{\rm GUT}$ in order not to spoil the gauge coupling
unification, they equally contribute to the anomaly
coefficients for SU(3)$_C$ and SU(2)$_L$. 
Therefore, the difference between these two anomaly coefficients 
${\cal A}_3$ and ${\cal A}_2$ is
independent of the choice of the extra matter multiplets, that is, 
the following condition has to be satisfied
\begin{eqnarray}
 {\cal A}_3 - {\cal A}_2\ =\ 2\alpha\ =\ \frac{N}{2}k''.
\end{eqnarray}
Since the integer $k''$ must be odd from Eq.(\ref{Masiero2'}),
it is found that $N=4M$. 

Obviously, the above extra matter multiplets are massless when the
discrete R-symmetry $Z_{NR}$ is exact. The Giudice-Masiero
mechanism seems the most natural way to induce their masses larger than 
the direct experimental bounds.
Therefore, the extra matter multiplets should satisfy the condition
\begin{eqnarray}
 x + \bar{x} = 0 \quad \mod N,
\label{extra}
\end{eqnarray}
where $x$ and $\bar{x}$ are the $Z_{NR}$ charge of the extra matter
multiplets $\Psi$ and $\bar{\Psi}$, respectively. 
For example, if we take one pair of ${\bf 5}$ and ${\bf 5^*}$ in the
SU(5)$_{\rm GUT}$ as the extra matter multiplets,
they contribute to the anomaly coefficients ${\cal A}_3$
and ${\cal A}_2$ by $-\alpha$.
Thus, it is found that this simplest choice is enough to make the
$Z_{NR}$ symmetry nonanomalous. 

By the same arguments as in the MSSM,
we can find that only $Z_{4R}$ and $Z_{20R}$ are independent 
and that the corresponding charge assignments are as follows:
\begin{eqnarray}
 Z_{4R} &:& P V^{2m} \quad (m=0,1)\\
 Z_{20R} &:& P^5 V^{2m} \quad (m=1,2,3,4,6,7,8,9)
\end{eqnarray} 
where the generators $P$ and $V$ are defined in Table \ref{generator2}.
$V^{2m}$ can be understood as a remaining discrete subgroup of U(1)$_V$
after generation of right-handed Majorana neutrino masses: U(1)$_V$
$\to$ $Z_{10V}$.
Therefore, if we gauge this U(1)$_V$ as in the SO(10)$_{\rm GUT}$,
the above ten discrete R-symmetries are equivalent to each other,
and the nonanomalous discrete R-symmetry is
uniquely determined as Z$_{4R}$ : $P$. 

\begin{table}
 \begin{center}
  \begin{tabular}{|c|rrrrrr|}
   \hline \raisebox{0ex}[12pt]{}
   \makebox[8mm]{} & 
   \makebox[5mm]{\hfill $t$} & \makebox[5mm]{\hfill $\bar{f}$} &
   \makebox[5mm]{\hfill $\bar{n}$} & \makebox[5mm]{\hfill $h$} &
   \makebox[5mm]{\hfill $\bar{h}$} & \makebox[5mm]{\hfill $\alpha$} \\ 
   \hline
   $P$ & $1$ &  $1$ & $1$ &  $0$ & $0$ & $1$ \\
   $V$ & $1$ & $-3$ & $5$ & $-2$ & $2$ & $0$ \\
   \hline
  \end{tabular}
  \caption{The charges of the two generators. $P$ is the same as in
  Table \ref{generator}. The generator $V$ corresponds to a U(1)
  subgroup of SO(10) ($\supset$ SU(5)$_{\rm GUT}\times$U(1)$_V$).}
  \label{generator2}
 \end{center}
\end{table}

It can be easily checked that these discrete R-symmetries forbid
all the unwanted dimension four and five operators. This is 
because an allowed operator has to satisfy 
the condition $Q_P-2m(Q_B-Q_L)=2\ (\mod 4)$, where $Q_L$ is 
the lepton number.
(Note that the generator $V$ is equivalent to $(B^{-1}L)^5$ up to the
U(1)$_Y$ gauge symmetry.)
After the SUSY breaking both $Z_{4R}$ and $Z_{20R}$ is broken down 
to $Z_{2R}$ (the R-parity), so that the unwanted dimension four operators
are forbidden completely.

We have embedded quarks and leptons in the GUT multiplets. 
As for the Higgs sector, however, a naive extension to the GUT
multiplets brings about massless colored Higgs multiplets.
The only model free from this problem
is the ``R-invariant natural unification'' model 
with a semisimple gauge group SU(5)$_{\rm GUT}\times$U(3)$_H$ 
\cite{yanagida,hy,iy}.
The matter content in this model is given in Table \ref{hyper},
where we take $Z_{4R}$ as the discrete R-symmetry for simplicity.
\begin{table}
 \begin{center}
  \begin{tabular}{|c|ccccc|ccccc|c|}
   \hline \raisebox{0ex}[11pt]{}
   & $T_{rs}$ & $\bar{F}^r$ & $\bar{N}$ & $H_r$ & $\bar{H}^r$ &
   $Q^r_\alpha$ & $\bar{Q}^\alpha_r$ &
   $Q^6_\alpha$ & $\bar{Q}^\alpha_6$ & $X^\alpha_\beta$ & $\theta$\\
   \hline
   SU(5)$_{\rm GUT}$ &
   ${\bf 10}$ & ${\bf 5^*}$ & ${\bf 1}$ & ${\bf 5}$ & ${\bf 5^*}$ &
   ${\bf 5^*}$ & ${\bf 5}$ & ${\bf 1}$ & ${\bf 1}$ & ${\bf 1}$ & \\
   \hline
   SU(3)$_H$ &
   ${\bf 1}$ & ${\bf 1}$ & ${\bf 1}$ & ${\bf 1}$ & ${\bf 1}$ &
   ${\bf 3}$ & ${\bf 3^*}$ & ${\bf 3}$ & ${\bf 3^*}$ & ${\bf 8}$ & \\ 
   \hline
   U(1)$_H$ &
   $0$ & $0$ & $0$ & $0$ & $0$ & 
   $1$ & $-1$ & $1$ & $-1$ & $0$ & \\ 
   \hline
   $Z_{4R}$ &
   $1$ & $1$ & $1$ & $0$ & $0$ &
   $0$ & $0$ & $2$ & $2$ & $2$ & $1$ \\
   \hline
  \end{tabular}
 \end{center}
 \caption{The matter content in the SU(5)$_{\rm GUT}\times$U(3)$_H$
 model.}
 \label{hyper}
\end{table}
The superpotential in this model is
\begin{eqnarray}
 W &=& Q^r_\alpha \bar{Q}^\beta_r X^\alpha_\beta 
  + Q^6_\alpha \bar{Q}^\beta_6 X^\alpha_\beta
  + Q^r_\alpha \bar{Q}^\alpha_6 H_r
  + \bar{Q}^\alpha_r Q^6_\alpha \bar{H}^r.
\end{eqnarray}
As shown in Ref.\cite{iy} 
only $Q^r_\alpha$ and $\bar{Q}_r^\alpha$ acquire the vacuum expectation
value $v$,
\begin{eqnarray}
 \vev{Q^r_\alpha} = v \delta^r_\alpha, \quad
 \vev{\bar{Q}_r^\alpha} = v \delta_r^\alpha.
\end{eqnarray}
Although $v$ is undetermined so far because of the presence of a flat
direction, it is not difficult to fix it to the GUT scale \cite{iy,ikny}.
In this vacuum, SU(3)$_C$ is an unbroken linear combination of an SU(3)
subgroup of the SU(5)$_{\rm GUT}$ and the hypercolor SU(3)$_H$, and
U(1)$_Y$ is that of a U(1) subgroup of the SU(5)$_{\rm GUT}$ and the
hyper U(1)$_H$.
The unification of the three gauge coupling constants of
SU(3)$_C$, SU(2)$_L$ and U(1)$_Y$ is practically achieved
when the SU(3)$_H$ and U(1)$_H$ gauge interactions are enough strong
at the GUT scale. 
The colored Higgs multiplets
acquire a mass of the order of the GUT scale
together with $Q^6_\alpha$ and $\bar{Q}_6^\alpha$.
On the other hand, the Higgs doublets do not have such a large mass 
because they have no partners to form $Z_{4R}$ invariant mass terms. 

One can easily check that the $Z_{4R}$ symmetry is free from the mixed
anomalies with SU(5)$_{\rm GUT}$ and SU(3)$_H$ in the presence of the
extra matter multiplets. It is also straightforward to check that the
mixed gravitational anomaly and the cubic anomaly of the $Z_{4R}$
symmetry are cancelled by adding some singlets.  
 
Finally, we comment on the extra matter multiplets.
For simplicity, we take one pair of $\Psi(\bf 5)$ 
and $\bar \Psi(\bf 5^*)$ as the extra matter multiplets. 
The sum of their charges is restricted by
Eq.(\ref{extra}):  $x+\bar{x}=0$, while each charge remains undetermined
from anomaly cancellation conditions. 
However, only the case that $x=-1$ and $\bar{x}=1$ 
turns out to be phenomenologically viable from the following arguments.

There can be four choices for the $Z_{4R}$ charge $x$ of
the extra matter multiplet $\Psi$, that is, $x=0,1,2,3$.\footnote{
There can be another choice that the extra matter multiplets 
have fractional charges. In this case the lightest extra
matter particle is stable. However, its relic abundance is 
not negligible, so that this choice is excluded
by the dark matter searches.
}
In the case $x=\bar{x}=0$, the extra matter multiplets have the same
quantum numbers as
the Higgs, allowing such a interaction as $TT\Psi$ and
$T\bar{F}\bar{\Psi}$.
Then the SU(3)$_C$ triplets of the extra matter multiplets
behave like the colored Higgs, and lead to too fast proton decays.
In the case $x=1$ and $\bar{x}=-1$,
one of the three matter multiplet $\bar{F}$ forms
a Planck mass term with the extra matter multiplets $\Psi$,
while the remaining $\bar{\Psi}$ cannot form a Yukawa interaction
$T\bar{\Psi}\bar{H}$. Therefore one of the down-type quarks and 
the charged leptons are massless. 
In the case that $x=2$ and $\bar{x}=-2$, 
the extra matter multiplets and the Higgs doublets
form Planck scale mass terms as $\Psi \bar{H}$ and $\bar{\Psi}H$,
so that no Higgs doublets remain at the weak scale. 
At last, it is found that the allowed choice is only that
$x=-1$ and $\bar{x}=1$.

Since the extra matter multiplets $\Psi$ and $\bar \Psi$
have the $Z_{4R}$ charges $x=-1$ and
$\bar{x}=1$ respectively, they have the following Yukawa couplings and
the mass terms\footnote{
Note that the third term $\vev{W}\Psi\bar{F}$ can be absorbed 
into the last term $\vev{W}\Psi\bar{\Psi}$ by 
an appropriate redefinition of the multiplets $\bar F$ and $\bar \Psi$.
}:
\begin{eqnarray}
 W&=& y T\bar{F}\bar{H} + z T\bar{\Psi}\bar{H} 
      + \vev{W}\Psi\bar{F} + \vev{W}\Psi\bar{\Psi}.  
\end{eqnarray}
Although there are mixings between the extra matter $\Psi$ and the 
matter multiplets $T$, 
these mixings are suppressed by $z\vev{\bar H}/\vev{W}$, 
so that their contributions to the FCNC processes are negligible. 
On the other hand, the Yukawa coupling $z T\bar{\Psi}\bar{H}$ 
could disturb the degeneracy among the masses of squarks and sleptons
in the matter multiplets $T$, 
whose effects on the FCNC processes could be observed by future
experiments.

\section{Conclusions}

In this paper we search for nonanomalous discrete R-symmetries
in the MSSM and the GUT from a phenomenological point of view. 
We find that the nonanomalous discrete R-symmetry should be
$Z_{9R}$ or $Z_{18R}$ in the MSSM.
In the case of the GUT, on the other hand, we find that there must be
some extra matter multiplets in order to cancel the anomalies and that
only the $Z_{4R}$ and $Z_{20R}$ symmetries are allowed.
The simplest choice of the extra matter multiplets is 
to take a pair of ${\bf 5}$ and ${\bf 5^*}$ of the SU(5)$_{\rm GUT}$.
These extra matter multiplets have masses of the order of the weak
scale, and the presence of them is testable in future hadron collider
experiments. As an example including the $Z_{4R}$ symmetry, we give a
unification model based on a semi-simple gauge group 
SU(5)$_{\rm GUT}\times$U(3)$_{\rm H}$. 

We conclude this paper with a comment on 
the next-to-minimal SUSY SM (NMSSM) \cite{NMSSM} 
with the gauge-mediated SUSY breaking (GMSB) \cite{GMSB}. 
In the GMSB, the NMSSM is an attractive mechanism 
generating the $\mu$-term because the Giudice-Masiero mechanism does not
work well. The Higgs sector in the NMSSM is $W=S
H\bar{H}+S^3$, where $S$ is a singlet field.
If the $Z_{NR}$ charge of $S$ is $2\alpha$,
the NMSSM is able to have the $Z_{4R}$ or $Z_{20R}$ symmetry.
In the NMSSM with the GMSB, as is known well, 
some extra fields are necessary to couple with the singlet $S$ in order
to obtain a viable spectrum \cite{GMSB,AG,GFM}.
It is very interesting that we have already introduced
such extra fields as the extra matter multiplets $\Psi$ and $\bar \Psi$
in order to cancel the anomalies of the  $Z_{4R}$ and $Z_{20R}$
symmetries.

\section*{Acknowledgements}

The authors would like to thank T. Watari for useful discussions.
This work was partially supported by the Japan Society for the
Promotion of Science (K.K. and N.M.) and ``Priority Area: Supersymmetry
and Unified Theory of Elementary Physics ({\#} 707)'' (T.Y.). 

%
%
%
\newcommand{\Journal}[4]{{\sl #1} {\bf #2} {(#3)} {#4}}
\newcommand{\PL}{\sl Phys. Lett.}
\newcommand{\PR}{\sl Phys. Rev.}
\newcommand{\PRL}{\sl Phys. Rev. Lett.}
\newcommand{\NP}{\sl Nucl. Phys.}
\newcommand{\ZP}{\sl Z. Phys.}
\newcommand{\PTP}{\sl Prog. Theor. Phys.}
\newcommand{\NC}{\sl Nuovo Cimento}
\newcommand{\MPL}{\sl Mod. Phys. Lett.}
\newcommand{\PRep}{\sl Phys. Rep.}

\end{document}